\documentclass{aa}  

\usepackage{graphicx}
\usepackage{txfonts}
\usepackage{booktabs} 
\usepackage{tikz} 
\usetikzlibrary{shapes.geometric, arrows}

\usepackage{siunitx}

\begin{document}

   \title{Applying Vision Transformers on Spectral Analysis of Astronomical Objects}
   
   \author{Luis Felipe Strano Moraes
          \inst{1}
          \and
          Ignacio Becker
          \inst{2}
          \and
          Pavlos Protopapas
          \inst{2}
          \and
          Guillermo Cabrera-Vives
          \inst{3,4,5,6,7}
          }

   \institute{Harvard Extension School, Harvard University, Cambridge, MA, 02138, USA
         \and
         John A. Paulson School of Engineering and Applied Science, Harvard University, Cambridge, MA, 02138, USA
         \and
         Department of Computer Science, Universidad de Concepción, Edmundo Larenas 219, Concepción, Chile \and
Center for Data and Artificial Intelligence, Universidad de Concepción, Edmundo Larenas 310, Concepción, Chile \and
Millennium Institute of Astrophysics (MAS), Nuncio Monseñor Sotero Sanz 100, Of. 104, Providencia, Santiago, Chile \and
Millennium Nucleus for Galaxies (MINGAL), Chile \and 
Heidelberg Institute for Theoretical Studies, Heidelberg, Baden-Württemberg, Germany}

   \date{Received May 30, 2025; accepted }

\abstract{
We apply pre-trained Vision Transformers (ViTs), originally developed for image recognition, to the analysis of astronomical spectral data. 
By converting traditional one-dimensional spectra into two-dimensional image representations, we enable ViTs to capture both local and global spectral features through spatial self-attention. 
We fine-tune a ViT pretrained on ImageNet using millions of spectra from the SDSS and LAMOST surveys, represented as spectral plots. 
Our model is evaluated on key tasks including stellar object classification and redshift (\( z \)) estimation, where it demonstrates strong performance and scalability. 
We achieve classification accuracy higher than Support Vector Machines and Random Forests, and attain \( R^2 \) values comparable to AstroCLIP's spectrum encoder, even when generalizing across diverse object types. 
These results demonstrate the effectiveness of using pretrained vision models for spectroscopic data analysis. 
To our knowledge, this is the first application of ViTs to large-scale, which also leverages real spectroscopic data and does not rely on synthetic inputs.
}


   \maketitle
%

\section{Introduction}

Spectroscopy is a core observational technique in astrophysics for determining the physical and chemical properties of celestial objects \citep{Burrows2010}. By dispersing light into a spectrum, astronomers can extract information about an object's composition, temperature, radial motion, and even aspects of its structure or environment \citep{Gray2005}. Unlike direct imaging, which mainly provides spatial information, spectroscopic observations probe the underlying physical processes and conditions in astronomical objects such as stars, nebulae, and galaxies. Through spectral analysis, scientists can identify the elements present in them and discern how they exist or interact under extreme cosmic conditions that cannot be replicated in laboratories \citep{Wahlgren2011}.

Spectral data are also essential for understanding the large-scale structure and evolution of the universe. The redshift of spectral lines provides a key method for measuring cosmic expansion, allowing us to estimate distances to galaxies and trace the large-scale structure of the cosmos \citep{hubble1929, Colless2001}. However, galaxy evolution is not solely dictated by their positions and motions in an expanding universe; their internal chemical composition also shapes it. Spectroscopy plays a crucial role in this aspect, revealing how elements are synthesized in stars, expelled into the interstellar medium, and recycled into subsequent generations of stars \citep{Maiolino2019}. By studying absorption and emission lines, astronomers can track the abundance of elements essential for planetary formation and, ultimately, for the emergence of life \citep{Wolfe2005}.

Modern spectroscopic surveys such as the Sloan Digital Sky Survey \citep[SDSS, ][]{kollmeier2019, Stoughton_2002} and the Large Sky Area Multi-Object Fiber Spectroscopic Telescope \citep[LAMOST, ][]{Luo_2015} have enabled access to datasets containing millions of spectra across diverse object types. These surveys are also making continuous data releases to the public, enabling groundbreaking research to take place. The volume of spectral data continues to grow, with surveys such as \citep[MSE, ][]{maunakea2023}, \citep[4MOST, ][]{4most2016}, \citep[GAIA, ][]{gaia2016}, and \citep[DESI, ][]{desi2023} now operational or in preparation.

Spectral redshift estimation and classification in surveys like SDSS typically rely on template-fitting, where observed spectra are matched to composite templates derived from empirically defined object classes \citep{kugler2015}. 
These templates are applied to each observed spectrum, allowing predefined properties such as the redshift to be computed by identifying the best fit. However, this approach simplifies the complexity of the data, limiting the precision of individual property estimates. Furthermore, the reliability of the results is sensitive to the selection and construction of the reference templates. While such automated pipelines improve efficiency over manual inspection, they still operate under constrained assumptions and leave room for more flexible, data-driven approaches that could capture finer-grained spectral features.

Manual inspection introduces uncertainty in redshift determination, often estimated as $\sigma_z / (1 + z) < 0.001$.
One possible cause of this uncertainty, as noted by \citet{yang2018}, is that up to $25\%$ of the input targets in their analysis could be categorized as unreliable due to low signal-to-noise ratios or ambiguous spectral features. So even introducing manual inspection still leaves possibilities for errors and with the size of the surveys, adding extra automated steps to help validate and cross-check means greater reliability on the results.

Deep learning models offer a promising alternative by learning generalizable spectral representations from datasets. Unlike manual inspection or traditional template-based approaches, these models can capture complex patterns in spectral data and generalize across different observational conditions. Vision Transformers \citep[ViTs]{dosovitskiy2020}, a specific Deep Learning architecture introduced for image recognition, are particularly promising for spectral analysis due to their ability to capture contextual information in a scalable manner that is also adaptable to multiple downstream tasks, making them suitable for a broad range of astrophysical experiments.

ViTs are derived from Transformer architectures, originally developed for natural language processing \citep{vaswani2017}, which leverage self-attention mechanisms to efficiently model long-range dependencies. Unlike recurrent architectures \citep{hochreiter1997}that process sequences sequentially, transformers simultaneously compute relationships between all tokens in parallel, enabling more efficient training and better scalability to long sequences and large datasets. The self-attention mechanism allows each token to assess the relevance of all other tokens, facilitating a contextual understanding that is particularly powerful in language tasks. Typically, transformers include a special classification (CLS) token, serving as a condensed representation of the entire sequence for downstream prediction tasks.

When adapted for images, ViTs divide the input image into fixed-size patches, which are then linearly embedded and combined with positional encodings to preserve spatial structure. These patch embeddings pass through transformer layers, where self-attention captures global interactions across the entire image. Unlike convolutional neural networks \citep[CNNs]{lecun1998gradient,krizhevsky2012imagenet}, which progressively build hierarchical features through local receptive fields, ViTs inherently model global relationships from the earliest layers. This capability motivates our approach to transform astronomical spectra into two-dimensional, image-like representations, enabling effective analysis using ViTs.

ViTs typically require very large training and finetuning datasets, a requirement that modern astronomy can meet with upcoming massive surveys. We employ state-of-the-art ViTs that were pre-trained on regular images from ImageNet \citep{deng2009imagenet} and finetune it on plots of spectra generated from a combined dataset comprising large portions of SDSS and LAMOST surveys.

We evaluate these models on multiple tasks, including redshift regression, stellar parameter inference, and morphological classification. In both types of tasks (classification and regression), the model shows high accuracy and performs well across a range of signal-to-noise (SNR) ratios. Furthermore, this design allows for easy integration of more data sources, such as different surveys, and refinements to the downstream tasks being performed. These results highlight the extensibility and the potential to support a broad range of future applications with our approach.

This paper is organized as follows. Section \ref{sec:prev} reviews related work. Section \ref{sec:data} describes the datasets used and their processing. Section \ref{sec:downstream} outlines the downstream tasks in detail. Section \ref{sec:model} discusses the model architecture. Section \ref{sec:results} presents results and we discuss them in Section \ref{sec:discussion}. Finally, Section \ref{sec:conclusion} concludes with future steps.

\section{Previous Work}\label{sec:prev}

Several traditional and automated approaches have been developed for redshift estimation using spectroscopic data. Among the most prominent is Redrock \citep{ross2020}, widely used in both the DESI project and recent SDSS data releases. Another notable method is \textsc{Darth Fader} \citep[DF]{machado2013}. Both techniques operate by cross-correlating observed spectra with templates over a range of redshift values and minimizing the $\chi^2$ error. These methods do not require prior knowledge of the physical properties of the sources and have demonstrated reasonable performance even under low SNR conditions.

In the last decade, efforts have incorporated machine learning techniques to improve redshift regression. \citet{fronterapons2019} introduced two models: one based on Dictionary Learning (DL) and another on a Denoising Autoencoder (DAE). The DL model learns a sparse dictionary of galaxy spectra and estimates redshift by minimizing the reconstruction error for a new input spectrum. The DAE model, on the other hand, is trained on synthetic galaxy spectra at zero redshift and estimates redshift by finding the transformation that best reconstructs the observed spectrum. A hybrid model that dynamically selects between DL and DAE based on input characteristics outperforms \textsc{Darth Fader} in comparable scenarios. Both \citet{machado2013} and \citet{fronterapons2019} focused exclusively on galaxies, and relied on simulated spectra.

More recently, \citet{podsztavek2022} proposed a redshift estimation framework using Bayesian convolutional neural networks, specifically designed to identify potentially unreliable redshift values in large spectroscopic surveys. Their architecture is based on the VGG network \citep{simonyan2015} and was trained on real spectroscopic data from \citet[SDSS Quasar Catalog DR12]{paris2017}. To evaluate performance, they also implemented a simpler Bayesian fully connected neural network (Bayesian FCNN) as a baseline. Their implementation treats the input spectrum as a 1D signal, mapping each wavelength to a single flux value, and does not leverage multiple color channels which are commonly used in standard image processing.

Beyond redshift estimation, other studies have focused on classification tasks using spectroscopic data. In this case, most studies apply dimensionality reduction techniques, such as Principal Component Analysis (PCA), followed by clustering or classification algorithms. For instance, \citet{marchetti2012} used PCA via Karhunen–Loève projections on galaxy spectra from the VIPERS survey \citep{scodeggio2018}. After reducing the dimensionality, they applied k-means clustering to group galaxies into early, intermediate, late, and starburst categories. Their method achieved results comparable to photometric approaches while leveraging the richer information content of spectra. However, it exhibited limitations, particularly in underrepresented classes such as active galactic nuclei (AGNs).

\citet{parker2024} introduce AstroCLIP, a cross-modal foundation model that jointly embeds galaxy images and spectra into a shared latent space through self-supervised transformer encoders aligned via contrastive learning. Their approach uses a ViT-based image encoder and a GPT-2-inspired \citep{radford2019language} spectral encoder adapted for masked modeling, where spectra are segmented and partially masked to encourage the model to capture meaningful spectral features without labeled data. AstroCLIP, trained on Dark Energy Spectroscopic Instrument \citep[DESI]{desi2023} data and Legacy Imaging Survey \citep[DESI-LS]{schlegel2021} imagery, outperforms supervised baselines on tasks like stellar mass and metallicity estimation, and significantly improves photometric redshift predictions compared to prior self-supervised methods.
Notably, aligning images and spectra helps the spectral embeddings organize more clearly around astrophysical properties, showing how multi-modal contrastive learning can outperform traditional single-modality methods.

Although our focus is on spectroscopy, several recent works in photometric classification are relevant for their methodological contributions.

At a coarse level of granularity, \citet{wang2022} performed Star–Galaxy–QSO classification for the J-PLUS survey \citep{cenarro2019} using photometric data. They trained multiple classifiers, including Support Vector Machines \citep[SVMs]{cortes1995svm} and Random Forests \citep[RFs]{breiman2001rf}, on data from SDSS, LAMOST, and the VERONCAT catalog \citep{veroncetty2010}. SVMs yielded the best performance, with RFs achieving nearly equivalent results.

Finer-grained photometric classification was explored in \citet{vavilova2021}, who benchmarked several methods and found SVMs and RFs to yield $96.4\%$ and $95.5\%$ accuracy, respectively. While CNNs performed slightly better (up to $98\%$), they claim they required high-resolution imaging and were less robust at higher redshifts.

\citet{daoutis2025} applied RFs to classify galaxies into star-forming, AGN, and passive categories, achieving around $99\%$ overall accuracy. Their model demonstrated particularly strong performance on star-forming galaxies, with slightly lower accuracy for AGNs.

Transformer-based architectures have also been explored in photometric contexts. \citet{donoso2023} introduced a Transformer model inspired by BERT \citep{devlin2018} for analyzing light curves, which they fine-tuned for both classification and regression tasks. Meanwhile, \citet{cao2024} combined CNNs with ViTs in a hybrid Convolutional Visual Transformer (CvT) architecture for galaxy morphology classification using image data from Galaxy Zoo \citep{willett2013}.

While these photometric studies do not operate on spectroscopic inputs, they highlight a broader interest in applying modern architectures, including Transformers and ViTs, to astronomical data, motivating our adaptation of ViTs for spectral analysis.

\section{Data Sources}\label{sec:data}

The training relies on spectroscopic data from two large-scale sky surveys: SDSS, specifically Data Release 18 and LAMOST, with Version 2.0 of Data Release 10. The diversity and volume of spectra from both surveys make them an ideal foundation for training a model capable of learning complex patterns and generalizing across varying observational conditions. In the following subsections, we describe each dataset’s characteristics, including their spectral coverage and selection criteria. For experimentation purposes, we split these datasets into: medium sized datasets which contain a balanced representation across classes, and big datasets which contain all of the objects in the each survey. Table~\ref{tab:dataset} summarizes the number of objects per morphological class for each dataset, and includes a new joint dataset combining Sloan and LAMOST which is referred to as SLOMOST, and served as the primary dataset for our experiments.

The wavelengths collected from the surveys span from about 3600 to 10400 \text{\AA}. The spectra obtained from SDSS has $R$ between $1560$ and $2650$, therefore we matched it by using the Low Resolution subset of the LAMOST survey which has $R \sim 1800$.

  \begin{table}
     \caption{Datasets and per-class representation. Values are in thousands.}
    \begin{tabular}{lrrrr}
      \toprule
Dataset & Stars & Quasars & Galaxies & Total\\
\midrule
      SDSS-Medium & 200 & 200 & 200 & 600 \\
      SDSS-Big & 322 & 668 & 1777 & 2767 \\
      LAMOST-Medium & 200 & 58 & 200 & 458 \\
      LAMOST-Big & 11013 & 58 & 236 & 11307 \\
      SLOMOST-Med & 400 & 258 & 400 & 1058 \\
      SLOMOST-Big & 11335 & 726 & 2013 & 14074 \\
      \bottomrule
    \end{tabular}

\label{tab:dataset}
  \end{table}

\begin{figure}[ht]
    \centering
	\fbox{\includegraphics[scale=0.4]{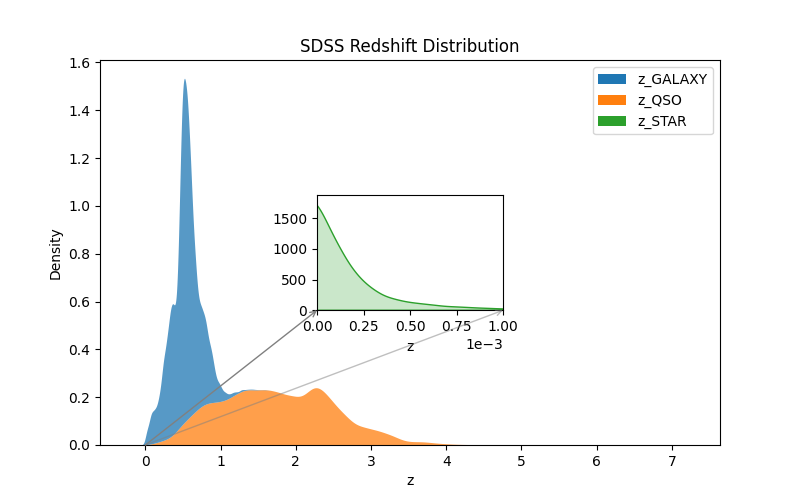}}
    \caption{Distribution of redshifts for the SDSS dataset}
    \label{fig:sdss-redshift}
\end{figure}

\begin{figure}[ht]
    \centering
	\fbox{\includegraphics[scale=0.4]{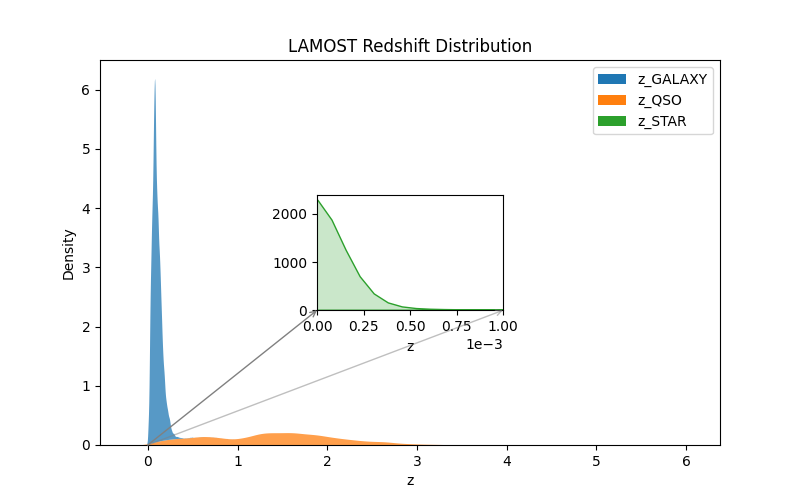}}
    \caption{Distribution of redshifts for the LAMOST dataset}
    \label{fig:lamost-redshift}
\end{figure}

\begin{figure}[ht]
    \centering
	\fbox{\includegraphics[scale=0.4]{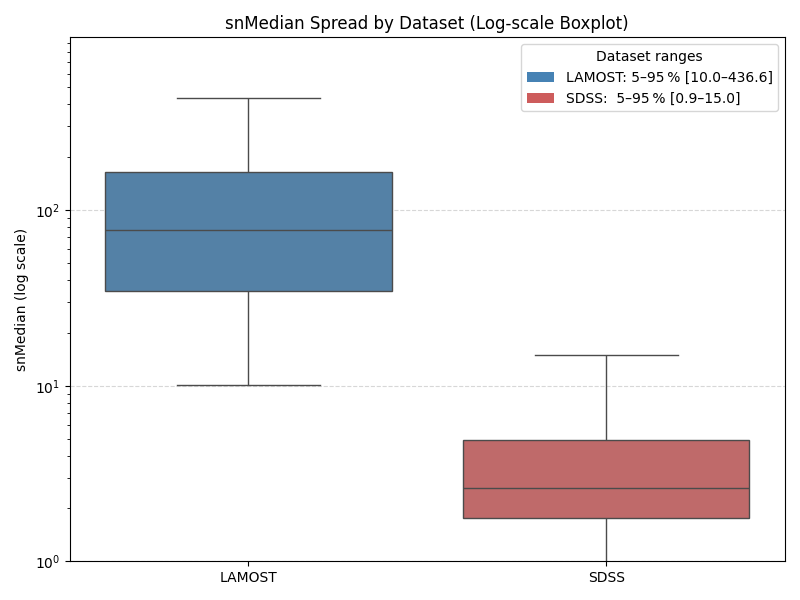}}
    \caption{Comparison of \texttt{snMedian} distributions in across both datasets via a log‑scale boxplot. The central box spans the interquartile range (25th–75th percentiles), whiskers extend to the 5th and 95th percentiles, and outliers beyond this range are omitted for clarity. }
    \label{fig:snmedian_sdss}
\end{figure}

Figures ~\ref{fig:sdss-redshift} and ~\ref{fig:lamost-redshift} shows the distribution of redshifts in both datasets across each of the major classes. Figure~\ref{fig:snmedian_sdss} displays the distribution of SNR values, using the \textit{snMedian} field provided by SDSS for each object\footnote{snMedian is a value provided by SDSS to represent an overall SNR value for the object across the different filter bands}. Since LAMOST does not provide a precalculated \textit{snMedian} value, we computed it based on SDSS conventions\footnote{We computed \textit{snMedian} as \(\textit{snMedian} = \sqrt{\sum_{i \in \{u, g, r, i, z\}} \textit{SNR}_i^2}\), following SDSS-derived practices.}.

\subsection{SDSS}\label{sec:SDSS}

SDSS, operational since 2000, has mapped millions of celestial objects---including stars, galaxies, and quasars---using fiber-optic spectrographs to capture optical and near-infrared data across a significant portion of the sky. It encompasses multiple spectroscopic programs targeting different object types. Notably, the BOSS (Baryon Oscillation Spectroscopic Survey) and eBOSS (Extended BOSS) components targeted galaxies up to $z \sim 1$ and quasars up to $z \sim 6$.

We excluded objects with $zWarning \neq 0$, or with $\text{instrument} \neq \text{'BOSS'}$ or $\text{targetType} \neq \text{'SCIENCE'}$. From an initial set of 5112k objects, this selection left us with approximately 2767k objects.

\subsection{LAMOST}\label{sec:lamost}

LAMOST, active since 2012, employs a wide-field design and fiber-optic technology to observe up to 4,000 objects simultaneously, focusing on stellar kinematics, chemical abundances, and radial velocities. It is optimized for high-throughput spectroscopic surveys of stars in the Milky Way, enabling large-scale studies of the structure, formation history, and kinematics of the Galaxy's disk and halo.

LAMOST primarily targets objects at \( z \approx 0 \) (Milky Way stars), with only a small fraction of low-redshift extragalactic sources.

We excluded entries where any of the fields \textit{z}, \textit{z\_err}, \textit{snru}, \textit{snrg}, \textit{snrr}, \textit{snri}, or \textit{snrz} were set to $-9999$, indicating data quality issues. From an initial set of 11441k objects, this left us with 11307 objects.

\section{Downstream Tasks}\label{sec:downstream}

\subsection{Stellar Object Classification}

The first downstream task we consider is the classification of astronomical sources. Traditionally, objects observed in spectroscopic surveys are broadly categorized into stars, galaxies, and quasars. These categories are central to astrophysical studies, enabling insights into stellar evolution, galactic structure, and accretion processes around supermassive black holes. We currently only implement and evaluate classification into broad categories, but the model is capable of finer subdivisions. These could provide astrophysical insights, for example stellar spectral types reveal temperature and composition, while galaxy subclasses indicate star formation rates or metallicity. Quasars may be further subdivided by emission line characteristics or luminosity classes, and galaxies can be categorized into morphological or spectroscopic subclasses that inform us about their star formation rates, dust content, and metallicity gradients.

Results for classification are shown in subsection \ref{sec:clasresults}, with a comparison of the different plot types in the table \ref{tab:perf-classification}.

\subsection{Redshift Regression}

The second key downstream task is the estimation of redshift for extragalactic objects.  The ViT-based architecture inherently captures global spectral patterns, making it well-suited to detect shifts in characteristic emission and absorption lines without being confounded by local noise or incomplete line profiles. By encoding an entire spectrum as a sequence of contextualized patches, the model can discern small wavelength shifts, even in the presence of multiple lines or low SNR ratios. As a result, the model produces accurate redshift estimates as seen in Tab. \ref{tab:perf-redshift} in subsection \ref{sec:redresults}.

\subsection{Stellar Parameter Regression}

Beyond redshift estimation, we also experiment with the regression of fundamental stellar parameters, including effective temperature ($T_{\text{eff}}$), surface gravity ($\log g$), and metallicity ($[\text{Fe/H}]$). The ability to infer these values directly from spectra allows for large-scale stellar population studies, aiding in Galactic archaeology and the study of stellar formation histories \citep{GaiaDR3_Apsis}. 

The results for each individual parameter can be found within subsection \ref{sec:steresults} below.

\section{Model Architecture}\label{sec:model}

\begin{figure}[ht]
    \centering
	\fbox{\includegraphics[scale=0.5]{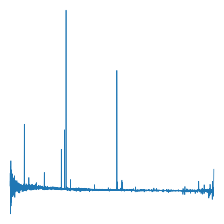}}
    \caption{\textit{Simple} plot type of spectra for one of the SDSS objects, a starbust Galaxy with ID 9068120565953615872}
    \label{fig:example_normal}
\end{figure}

We represent the spectral data visually, based on the hypothesis that image-based formats may reveal patterns more readily learnable by the model. Figure ~\ref{fig:example_normal} shows an example of a galaxy spectra that was used as input to the model during  experiments. Although it is a direct plot of the spectral values, experiments with this representation already delivered good performance, and for regression of effective temperature and surface gravity of stars, it proved to be the most effective.
We have also explored other representations which either modified the format of how the information was embedded into the final input as well as attempts with adding extra information. In the following sections, we discuss the various processing methods that led to improved performance on each specific task. For consistency and ease of comparison, all subsequent visualizations are based on the same galaxy.

\begin{figure}[ht]
    \centering
	\fbox{\includegraphics[scale=0.22]{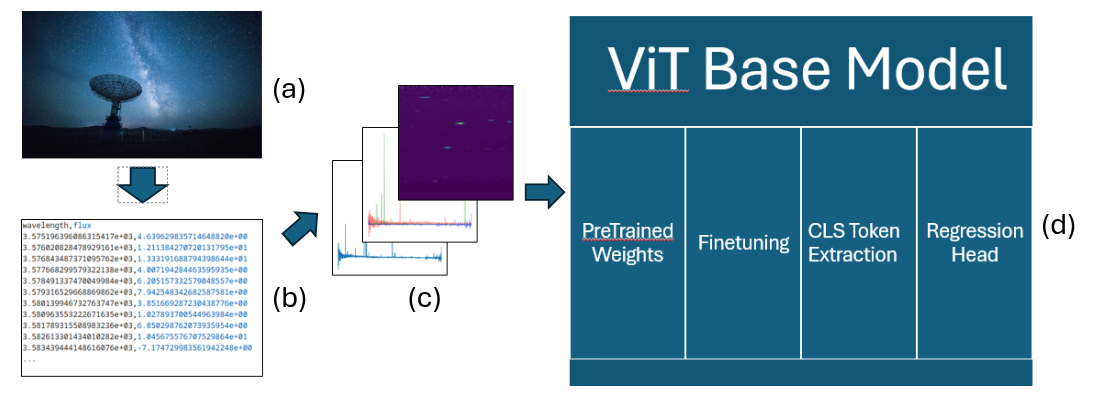}}
    \caption{Model pipeline example for a regression task: (a) data is obtained from surveys (b) processed and kept in local files (c) goes through generation of different plot types (d) passes through the ViT Base Model for finetuning.}
    \label{fig:diagram}
\end{figure}

Our underlying model is based on \textsc{DINO} from \citet{dino2021}, a self-supervised learning approach described as a form of self-distillation without labels. The model was pretrained on the ImageNet dataset in an unsupervised manner. We used the pretrained version of DINO as a backbone for our models without pretraining it for spectral data.

For the redshift regression task, we finetune a pretrained ViT model obtained from Hugging Face (namely \textit{facebook/dino-vitb16}), which processes input images resized to $224\times224$ pixels, with a custom regression head. This head consists of a single linear layer that maps the CLS token output from the ViT to a single continuous value representing the predicted redshift. During training, we use Mean Squared Error (MSE) as the loss function. The input images are normalized using a mean and standard deviation of 0.5 for each color channel.

\subsection{Pipeline}
The pipeline, shown in Figure \ref{fig:diagram} is a structured workflow that converts raw spectroscopic data into formats optimized for analysis with Vision Transformers (ViTs). This section outlines each stage of the pipeline, from data acquisition through pre-processing to model pre-training and fine-tuning.

The pipeline begins by acquiring spectroscopic data and metadata from publicly available surveys such as SDSS and LAMOST. These datasets, typically in FITS format, are processed using the \texttt{AstroPy} library. Each object is then saved as an individual CSV file containing wavelength and flux columns, simplifying downstream processing and model input preparation.

Next, several preprocessing steps are applied to ensure data quality and consistency: the wavelengths are normalized to standardize their representation across different observations, the flux values are normalized to mitigate variations due to differences in instrument sensitivity or observational conditions, and we perform some control of outlier values by setting up thresholds based on the first and last quartile of the wavelengths from the dataset.

To adapt spectral data for consumption by Vision Transformers (ViTs), one-dimensional spectra are transformed into two-dimensional image representations. This transformation is a key component of our model design, as the choice of how the spectra are encoded into images can significantly influence performance.  Initial experiments used a plot we refer to as \textit{Simple}, where the original spectrum is simply drawn as in Figure~\ref{fig:example_normal}. We then explored alternative encodings, with results presented below. In Figure~\ref{fig:example_overlap}, we present a plot called \textit{Overlap}, where we divide the spectrum into three segments of equal length and map each to a separate RGB color channel to determine if this approach yields any performance gains, due to allowing for larger detail of each section of the spectra to be shown in the final image. 

\begin{figure}[ht]
    \centering
	\fbox{\includegraphics[scale=0.5]{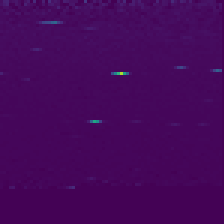}}
    \caption{\textit{2D Map} plot type of the spectra, where we associate each individual wavelength with a square of 3x3 pixels in the final plot}
    \label{fig:example_map2d}
\end{figure}

\begin{figure}[ht]
    \centering
	\fbox{\includegraphics[scale=0.3]{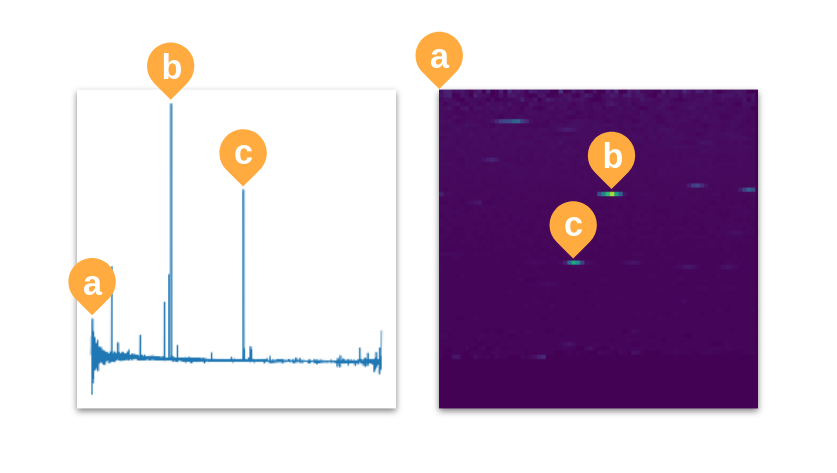}}
    \caption{Overview of how each individual flux is mapped to the final 2D image in the 2D Map design. Labels $a$, $b$ and $c$ can be seen on the left in the standard flux plot, and in the right side with intensity set as the color of a given region in the image}
    \label{fig:map2d-mapping}
\end{figure}

Since the usual plots tend to have mostly empty space, we further introduce a more information-dense approach, which we call \textit{2D Map}, which replaces the standard line plot with a heatmap-like representation. We convert the one-dimensional flux data into a two-dimensional image representation by reshaping them into a square image of dimensions $224 \times 224$ pixels. This reshaping is executed by populating the image in fixed-size blocks of $3 \times 3$ pixels, where each flux value fills an individual block uniformly. Sequential spectral data points are thus systematically arranged into this spatial grid, visually encoding the spectral features. After this block-filling procedure, the method applies a colormap to the resulting 2D array based on the intensity of the flux values. The flux values are normalized between predefined minimum and maximum flux thresholds, ensuring consistent visual representation across different spectral data sets. The image is generated without axes and margins to create a clear and concise visualization. The resulting visualization is shown in Figure~\ref{fig:example_map2d}, and the mapping process is illustrated in detail in Figure~\ref{fig:map2d-mapping}.

\begin{figure}[ht]
    \centering
	\fbox{\includegraphics[scale=0.5]{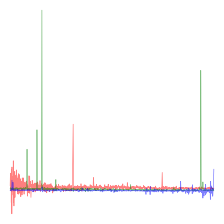}}
    \caption{\textit{Overlap} plot type for spectra where each color channel of an RGB image contains one third of the data}
    \label{fig:example_overlap}
\end{figure}

The final step of the pipeline involves using the generated 2D spectral images to fine-tune the base ViT model which was pretrained on large-scale image datasets to adapt its weights to spectral data. The fine-tuning step involves training the model on the spectral image dataset using a lower learning rate, allowing it to specialize in recognizing spectral features while retaining general representations learned during pretraining. During fine-tuning, task-specific heads are added to the ViT for classification (e.g., star/galaxy/quasar identification) or regression (e.g., redshift estimation). The performance of the fine-tuned model is validated using standard evaluation metrics such as accuracy, $F_1$-score, and MSE.

\section{Results}\label{sec:results}

In this section, we present the outcomes of our experiments and analyses. We begin by describing the evaluation metrics and datasets used for benchmarking. We then provide qualitative and quantitative results for both the classification and regression tasks. Finally, we compare our model’s performance against established baselines and discuss the implications of these findings.

The reported results are based on models finetuned on either the SLOMOST-Med or SLOMOST-Big datasets, as indicated in the description of each table. All models were finetuned for at least 30 epochs, with the best-performing checkpoint saved and used for evaluation. Hyperparameter tuning was conducted, and only the best results are shown. The optimal hyperparameters, a weight decay of $0.01$ and a learning rate of $10^{-5}$, were selected based on the tuning results. 

\subsection{Classification Results}\label{sec:clasresults}

Classification results are presented in Table~\ref{tab:perf-classification} with overall accuracy and macro-averaged F1 scores for each type of input image. Table~\ref{tab:recall-classification} shows per-class recall for each of these same image types. The finetuning was performed by minimizing the categorical cross entropy loss. For the \textit{2D Map} representation, which achieved the best performance across the variants when finetuning on SLOMOST-Med, we also report results for SLOMOST-Big, and present a Confusion Matrix in Table ~\ref{tab:conf-2dmap-classification-big}. 

\begin{table}[h!]
\centering
\begin{tabular}{ccc}
\hline
	\textbf{Image Type} & \textbf{Accuracy} & \textbf{F1}\\ \hline
	\textit{Simple} Plot SLOMOST-Med & 0.982 & 0.975\\ 
	\textit{Overlap} Plot SLOMOST-Med & 0.982 & 0.975\\ 
	\textit{2D Map} SLOMOST-Med & 0.990 & 0.991\\ 
   	\textit{2D Map} SLOMOST-Big & 0.994 & 0.991\\ \hline
\end{tabular}
\caption{Performance results for the different image types in doing morphological classification}
\label{tab:perf-classification}
\end{table}

\begin{table}[h!]
\centering
\begin{tabular}{cccc}
\hline
	\textbf{Image Type} & \textbf{Galaxy} & \textbf{QSO} & \textbf{Star}\\ \hline
	\textit{Simple} Plot SLOMOST-Med & 0.989 & 0.977 & 0.957\\ 
	\textit{Overlap} Plot SLOMOST-Med & 0.991 & 0.975 & 0.959\\ 
	\textit{2D Map} SLOMOST-Med & 0.992 & 0.988 & 0.993\\ 
   	\textit{2D Map} SLOMOST-Big & 0.997 & 0.986 & 0.990\\ \hline
\end{tabular}
\caption{Per class recall for each different image type in doing morphological classification}
\label{tab:recall-classification}
\end{table}

\begin{table}[h!]
\centering
\begin{tabular}{lccc}
\toprule
\textbf{Confusion Matrix} & \textbf{GALAXY} & \textbf{QSO} & \textbf{STAR} \\ \midrule
\textbf{GALAXY}           & 353421           & 1114          & 38            \\
\textbf{QSO}              & 1785             & 130124        & 26           \\
\textbf{STAR}             & 32              & 28          & 5766          \\ \bottomrule
\end{tabular}
\caption{\textit{2D Map} Plot Confusion Matrix: Predicted vs. True Labels (SLOMOST-Big)}
\label{tab:conf-2dmap-classification-big}
\end{table}

\subsection{Redshift Estimation Results}\label{sec:redresults}

\begin{figure}[ht]
    \centering
	\fbox{\includegraphics[scale=0.3]{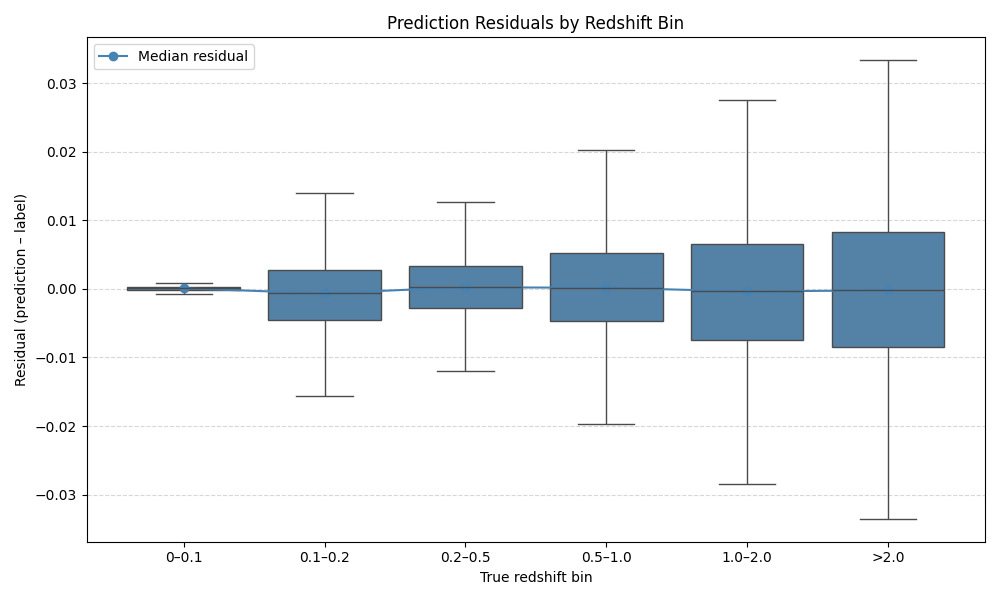}}
    \caption{Residuals of model predictions displayed as boxplots across true redshift bins. Each box spans the interquartile range (25th–75th percentiles) of the residual distribution within that bin, whiskers extend to the 5th and 95th percentiles. Prediction results from regression with \textit{2D Map} over SLOMOST-Big}
    \label{fig:reg-residuals}
\end{figure}

We present analogous results for the redshift regression task. Table~\ref{tab:perf-redshift} reports the $R^2$ scores for each input image type, including results for the \textit{2D Map} representation trained on SLOMOST-Big. Figure~\ref{fig:reg-residuals} shows the plot of predicted versus true redshift with \textit{2D Map} when finetuning with SLOMOST-Big.

Table~\ref{tab:snr-redshift} summarizes the model performance in various SNR bins. Following the evaluation criteria proposed in \cite{ross2020}, we define a non–catastrophic redshift estimation as one in which the difference between predicted and true redshift corresponds to a velocity offset smaller than $3000 \text{km s}^{-1}$ for quasars, and $1000 \text{km s}^{-1}$ for galaxies and stars, that is, where $\Delta z$, the absolute difference between predicted and true redshift, remains below these thresholds. Performance starts to drop in higher SNR bins, but that matches a significant drop in the amount of objects available for evaluation in them (e.g. only $0.02\%$ of objects fall in the $> 50$ SNR bin).

\begin{table}[h!]
\centering
\caption{Performance results for the different image types in doing redshift regression. }
\begin{tabular}{cc}
\hline
\textbf{Image Type} & \textbf{R2} \\ \hline
	\textit{Simple} Plot SLOMOST-Med & 0.942 \\ 
	\textit{Overlap} Plot SLOMOST-Med & 0.939 \\ 
	\textit{2D Map} SLOMOST-Med & 0.980\\ 
    \textit{2D Map} SLOMOST-Big & 0.992 \\ \hline
\end{tabular}

\label{tab:perf-redshift}
\end{table}

\begin{table}[h!]
\centering
\begin{tabular}{cc}
\hline
\textbf{SNR Range} & \textbf{Success} \\ \hline
0-1 & 59.01 \\
1-2 & 69.05 \\
2-5 & 80.19 \\
5-7 & 85.75 \\
7-10 & 86.54 \\
10-20 & 85.76 \\
20-30 & 82.11 \\
30-40 & 72.98 \\
40-50 & 64.00 \\
50+ & 59.80 \\
\end{tabular}
\caption{Performance of redshift regression over different SNR ranges with \textit{2D Map}}
\label{tab:snr-redshift}
\end{table}

\subsection{Results on stellar parameters}\label{sec:steresults}

Finally, we report $R^2$ for three stellar parameter regression tasks. Table ~\ref{tab:perf-teff} and Table ~\ref{tab:perf-logg} present results for $T_{\mathrm{eff}}$ and $\log g$, respectively,  where the \textit{Simple} plot yielded the best performance. Table~\ref{tab:perf-feh} reports the $[\text{Fe/H}]$ regression results, for which the \textit{Overlap} plot provided superior outcomes.

\begin{table}[h!]
\centering
\begin{tabular}{cc}
\hline
\textbf{Image Type} & \textbf{R2} \\ \hline
	\textit{Simple} Plot SLOMOST-Med & 0.739 \\
	\textit{Overlap} Plot SLOMOST-Med & 0.728 \\
	\textit{2D Map} SLOMOST-Med & 0.664 \\
    \textit{Simple} Plot SLOMOST-Big & 0.799 \\
\end{tabular}
\caption{Performance results for the different image types in doing effective temperature regression}
\label{tab:perf-teff}
\end{table}

\begin{table}[h!]
\centering
\begin{tabular}{cc}
\hline
\textbf{Image Type} & \textbf{R2} \\ \hline
	\textit{Simple} Plot SLOMOST-Med & 0.695 \\
	\textit{Overlap} Plot SLOMOST-Med & 0.682 \\
	\textit{2D Map} SLOMOST-Med & 0.637 \\
    \textit{Simple} Plot SLOMOST-Big & 0.790 \\
\end{tabular}
\caption{Performance results for the different image types in doing surface gravity regression}
\label{tab:perf-logg}
\end{table}

\begin{table}[h!]
\centering
\begin{tabular}{cc}
\hline
\textbf{Image Type} & \textbf{R2} \\ \hline
	\textit{Simple} Plot SLOMOST-Med & 0.415 \\
	\textit{Overlap} Plot SLOMOST-Med & 0.427 \\
	\textit{2D Map} SLOMOST-Med & 0.324 \\
    \textit{Overlap} Plot SLOMOST-Big & 0.780 \\
\end{tabular}
\caption{Performance results for the different image types in doing metallicity regression}
\label{tab:perf-feh}
\end{table}

\section{Discussion}\label{sec:discussion}

Our experiments show that converting astronomical spectra into two-dimensional image formats enables Vision Transformers to effectively capture both global and local spectral features. Among the representations explored, the \textit{2D map} format consistently demonstrated strong performance across tasks, although it did not outperform all alternatives in every setting as can be seen in the previous section.

Table~\ref{tab:perf-vsbsznet} compares the performance of redshift regression in our model against the Bayesian SZNet model introduced by \citet{podsztavek2022}, as well as a simpler baseline model they implemented, Bayesian FCNN. The results for Bayesian SZNet are taken directly from their publication and are based on spectra exclusively from quasars in the SDSS DR 12. For our work, we report results both on the quasar-only subset and on the full test set, which includes objects from both SDSS and LAMOST, using the test portion of the SLOMOST-Big dataset. 
To enable comparison with \citet{podsztavek2022}, who report only the Continuous Ranked Probability Score (CRPS), we computed CRPS under a Gaussian assumption: the predicted values were treated as the means of Gaussian distributions, and the standard deviation was estimated from the residuals between predictions and ground truth labels.

\begin{table}[h!]
\centering
\begin{tabular}{ccc}
\hline
\textbf{Model} & \textbf{RMSE} & \textbf{CRPS} \\ \hline
Bayesian SZNet (DR12Q) & 0.1083 & 0.0171 \\
Bayesian FCNN (DR12Q) & 0.2106 & 0.0712 \\
\textit{2D Map SLOMOST-Big} (Quasar only) & 0.1546 & 0.0277 \\
\textit{2D Map SLOMOST-Big} & 0.0397 & 0.0108 \\
\end{tabular}
\caption{Redshift regression performance of \textit{2D Map} vs Bayesian SZNet}
\label{tab:perf-vsbsznet}
\end{table}

Compared to AstroCLIP, which reports an $R^2$ of 0.990 for redshift regression using its spectrum encoder, we achieve a similar performance with an $R^2$ of 0.992. Notably, our model was trained and evaluated on a substantially larger and more diverse dataset. The AstroCLIP encoder was trained over 500 epochs in a single day on approximately 200,000 spectra using four NVIDIA H100 GPUs. In contrast, we performed finetuning over nine days on a single NVIDIA 4090 GPU, using spectra from more than $14M$ objects.

For classification, we compare our results to those of the SVM classifier from \citet{wang2022}, which was trained on a dataset that partially overlaps with the one used by us. Table~\ref{tab:perf-vsjplus} reports the accuracy of our model on our test set, alongside the accuracy of their SVM model and several other classification approaches evaluated in their study.
\begin{table}[h!]
\centering
\begin{tabular}{cc}
\hline
\textbf{Model} & \textbf{Accuracy} \\ \hline
Decision Tree & 92.6\% \\
Linear Discrimination & 86.9\% \\
Bayesian & 74.3\% \\
SVM & 96.4\% \\
k-NN & 95.7\% \\
AdaBoost & 92.0\% \\
Random Forest & 96.2\% \\
\textit{2D Map} & 99.0\% \\
\end{tabular}
\caption{Classification accuracy of multiple models from Wang et al. 2022 versus \textit{2D Map}}
\label{tab:perf-vsjplus}
\end{table}

In the classification task, our model achieved near-perfect performance, reaching $99.0\%$ accuracy on the SLOMOST-Med and SLOMOST-Big datasets when using the \textit{2D map} representation. Confusion matrices reveal that this representation particularly reduces misclassifications between quasars and galaxies, a category where less processed formats such as the \textit{Simple} and \textit{Overlap} plots exhibited greater confusion. We attribute this improvement to \textit{2D map}'s capacity to more clearly emphasize spatial differences between emission and absorption line features that are essential for distinguishing among object types.

For redshift regression, the Vision Transformer architecture benefited from its ability to model long-range dependencies within spectral structure. We achieved an $R^2$ score of 0.992 on the diverse SLOMOST-Big dataset, indicating strong generalization across multiple source types, spectral ranges, and observational conditions. The model also maintained high accuracy across a wide range of SNR ratios, with the best performance occurring at intermediate SNR values. The decline in performance at very high SNR levels may reflect limited sample sizes in those bins, although this trend merits further investigation.

Regression of stellar parameters lower $R^2$ values than redshift regression but still meaningful. The \textit{Simple} and \textit{Overlap} representations often yielded better performance than the \textit{2D map}, particularly for $T_{\text{eff}}$ and $\log g$. The relatively lower performance of the \textit{2D map} in these cases may indicate challenges in encoding subtle spectral features that influence these parameters. These observations suggest that targeted architectural adaptations, such as attention mechanisms incorporating spectral priors or hybrid image-sequence models, may be beneficial for improving performance in stellar parameter regression tasks.

In comparisons with baseline models, our model demonstrated competitive or superior performance as seen in the previous sections. Though we were not able to reproduce the exact results from the other studies given availability and ease of reproduction, we have used similar datasets and made observations of where they differ. For redshift regression, it outperformed Bayesian SZNet on diverse test sets, while achieving similar CRPS scores on quasar-only data. In classification, it surpassed several conventional machine learning approaches, including SVMs and RFs, by achieving higher accuracy on overlapping datasets. Notably, these results were obtained using a single ViT-based architecture with minimal task-specific tuning, which supports the potential of this work as a general-purpose framework for spectral analysis.

In summary, the results highlight both the strengths and limitations of applying visual transformer-based models to spectroscopic data. While we achieve strong results in classification and redshift estimation, stellar parameter regression presents additional challenges that may require further methodological refinement. Future work could explore the integration of domain-specific knowledge into model architectures or the use of multi-modal inputs that combine image representations with tabular or sequence-based data. As spectroscopic surveys continue to scale, newer models based on ViTs offer a promising foundation for efficient and accurate analysis of large volumes of spectral data.

\section{Conclusion and Future Work}\label{sec:conclusion}

In this paper, we applied Vision Transformers to perform several tasks in astronomy, leveraging a novel framework for analyzing astronomical spectral data using pre-trained ViTs. By transforming one-dimensional spectra into two-dimensional image representations and leveraging pretrained ViT backbones, we demonstrated that this approach can effectively capture complex spectral features and yield strong performance in both classification and regression tasks, offering a flexible and modular foundation for further experimentation and downstream applications in spectral analysis.

Our results highlight the potential for adapting modern deep learning architectures to the challenges of astrophysical data, where the volume, heterogeneity, and complexity of observations continue to grow. By bridging the gap between traditional spectral formats and visual transformer-based models, this work enables a more scalable and accurate analysis of large datasets. This work represents a step toward the development of general-purpose tools for spectral science, contributing to the broader goal of advancing our understanding of the composition, structure, and evolution of the universe.

\subsection{Next Steps}

Looking ahead, several directions for extension and refinement are available. These can be grouped into the following categories:

\begin{itemize}
    \item \textbf{Model improvements:} Future work includes incorporating additional spectral datasets from surveys not yet covered in this study, exploring alternative image representations, and experimenting with enriched visual encodings. For example, multichannel spectral plots inspired by the \textit{Overlap} format could include first derivatives, continuum-subtracted flux, or line detection maps. In addition, we plan to explore multimodal architectures, such as those inspired by AstroCLIP, that combine spectral data with complementary metadata or photometric information.
    
    \item \textbf{Extended downstream tasks:} Beyond coarse classification and redshift estimation, future work could involve fine-grained stellar or galaxy subclassification, as well as anomaly detection tasks for identifying rare or unusual spectral types. These directions will require the model to learn more nuanced spectral cues and may benefit from specialized loss functions or attention mechanisms.
    
    \item \textbf{Architectural refinements:} Further experiments are planned to evaluate changes in the ViT architecture itself, including variations in patch size, transformer depth, and token pooling strategies. Although initial attempts at pretraining ViTs from scratch on spectroscopic data did not yield notable gains, more targeted pretraining or contrastive learning approaches may improve generalization. 
    
 \item \textbf{Cross-domain applications:} We also intend to explore applications of this framework outside of astronomy. In fields such as agriculture, environmental monitoring, or materials science, spectral measurements are commonly used for classification or regression tasks. Our domain-agnostic structure and modular pipeline make it a promising candidate for transfer to these domains, provided appropriate training data are available.
\end{itemize}

Overall, this work lays the groundwork for more flexible, accurate, and scalable spectral analysis pipelines. We hope it contributes to the development of new machine learning techniques and practical tools for the next generation of spectroscopic surveys and beyond.

All of the source code for these experiments can be found at \url{https://github.com/astromer-science/spectromer} and we welcome contributions and feedback. Metadata files for both of the surveys used here are also provided, as well as documentation on how to introduce data from a new survey.

\begin{acknowledgements}

Funding for the Sloan Digital Sky Survey V has been provided by the Alfred P. Sloan Foundation, the Heising-Simons Foundation, the National Science Foundation, and the Participating Institutions. SDSS acknowledges support and resources from the Center for High-Performance Computing at the University of Utah. SDSS telescopes are located at Apache Point Observatory, funded by the Astrophysical Research Consortium and operated by New Mexico State University, and at Las Campanas Observatory, operated by the Carnegie Institution for Science. The SDSS web site is \url{www.sdss.org}.

SDSS is managed by the Astrophysical Research Consortium for the Participating Institutions of the SDSS Collaboration, including Caltech, The Carnegie Institution for Science, Chilean National Time Allocation Committee (CNTAC) ratified researchers, The Flatiron Institute, the Gotham Participation Group, Harvard University, Heidelberg University, The Johns Hopkins University, L’Ecole polytechnique f\'{e}d\'{e}rale de Lausanne (EPFL), Leibniz-Institut f\"{u}r Astrophysik Potsdam (AIP), Max-Planck-Institut f\"{u}r Astronomie (MPIA Heidelberg), Max-Planck-Institut f\"{u}r Extraterrestrische Physik (MPE), Nanjing University, National Astronomical Observatories of China (NAOC), New Mexico State University, The Ohio State University, Pennsylvania State University, Smithsonian Astrophysical Observatory, Space Telescope Science Institute (STScI), the Stellar Astrophysics Participation Group, Universidad Nacional Aut\'{o}noma de M\'{e}xico, University of Arizona, University of Colorado Boulder, University of Illinois at Urbana-Champaign, University of Toronto, University of Utah, University of Virginia, Yale University, and Yunnan University.

Guoshoujing Telescope (the Large Sky Area Multi-Object Fiber Spectroscopic Telescope LAMOST) is a National Major Scientific Project built by the Chinese Academy of Sciences. Funding for the project has been provided by the National Development and Reform Commission. LAMOST is operated and managed by the National Astronomical Observatories, Chinese Academy of Sciences.

GCV acknowledges support from the National Agency for Research and Development (ANID) grants: Millennium Science Initiative ICN12\_009, AIM23-0001, NCN2021\_080, NCN2024\_112, and FONDECYT Regular 1231877.

\end{acknowledgements}
\bibliographystyle{aa} 

\bibliography{article} 

\end{document}